\documentclass[USenglish]{article}

\usepackage[utf8]{inputenc}
\usepackage{natbib}
\usepackage[small]{dgruyter}
\usepackage{microtype}
\usepackage{color}

\setcitestyle{authoryear,open={(},close={)}}

\begin{document}

% \articletype{Article}
\newif\ifarxiv
\arxivtrue
% \arxivfalse 

\runningauthor{Nakahara et al.}
\title{Pitching strategy evaluation via stratified analysis using propensity score}
  
\author[1]{Hiroshi Nakahara}
\author[2]{Kazuya Takeda}
\author*[3]{Keisuke Fujii}
\affil[1]{Graduate School of Informatics, Nagoya University, Email: nakahara.hiroshi@g.sp.m.is.nagoya-u.ac.jp}
\affil[2]{Graduate School of Informatics, Nagoya University, Email: kazuya.takeda@nagoya-u.jp}
\affil[3]{Graduate School of Informatics, Nagoya University; RIKEN Center for Advanced Intelligence Project; PRESTO, Japan Science and Technology Agency, Email: fujii@i.nagoya-u.ac.jp}
\runningtitle{Pitching strategy evaluation via stratified analysis}
\subtitle{}
% abstract: 200 words
\abstract{Recent measurement technologies enable us to analyze baseball at higher levels. 
There are, however, still many unclear points around the pitching strategy.
The two elements make it difficult to measure the effect of pitching strategy.
First, most public datasets do not include location data where the catcher demands a ball, which is essential information to obtain the battery's intent.
Second, there are many confounders associated with pitching/batting results when evaluating pitching strategy.
We here clarify the effect of pitching attempts to a specific location, e.g., inside or outside.
We employ a causal inference framework called stratified analysis using a propensity score to evaluate the effects while removing the effect of disturbing factors.
We used a pitch-by-pitch dataset of Japanese professional baseball games held in 2014-2019, which includes location data where the catcher demands a ball.
The results reveal that an outside pitching attempt is more effective than an inside one to minimize allowed run on average.
Besides, the stratified analysis shows that the outside pitching attempt was always effective despite the magnitude of the \textcolor{black}{estimated} batter's ability, and the ratio of pitched inside for pitcher/batter.
Our analysis would provide practical insights into selecting a pitching strategy to minimize allowed runs.}
  \keywords{baseball; pitching strategy; causal inference; propensity score}
  \startpage{1}
  \aop

\maketitle
%%%%%%%%%%%%%%%%%%%%%%%%%%%%%%%
\vspace{-0pt}
\section{Introduction}
\vspace{-0 pt}

% 野球とデータ分析
Baseball is a good friend of statistics due to its features.
Each scene of baseball is discrete, making it easy to allocate responsibilities of each play to players.
Many formulas and statistics (stats) were developed and employed to evaluate players' performance from a long time ago \citep{james10, lewis04, click06, tango07, beneventano12, costa12}.
Most stats are computed using batting, pitching, and fielding results (for detail, see Section \ref{sec:related}).
Currently, measurement technologies and data platforms are getting developed, and
all sensing data measured in Major League Baseball (MLB) games are provided and accessible.
% It enables us to develop stats to estimate player's performance more accurately computed by speed and angle of batted balls, such as expected weighted On Base Average (xwOBA) and expected Earned Run Average (xERA) \citep{xwoba, xera}, 
\textcolor{black}{It enables us to develop stats to evaluate player's performance using speed and angle of batted balls, such as expected weighted On Base Average (xwOBA) and expected Earned Run Average (xERA) \citep{xwoba, xera}, 
which can acquire more important information and findings to win.}
``Fly ball revolution'', which specifies the effective speed and angle of batted balls to hit a long hit, is a symbol of that.
% The specific value of batted ball speed and angle to hit a home run (or long hit) were clarified

% 野球での配球の説明、位置付け
% In this research, we discuss about pitching strategy. 
% (i.e. which type and course should they pitch to minimize allowed runs)
% Pitching strategy means (here) strategy of battery (pitcher and batter) to defeat opponent batter.
% Battery combines various pitch type, course and pitch combination (e.g., inside fastball after outside breaking ball).
% During game, determining next pitch type/course is catcher's task basically.
% He determines next one and send a signal to pitcher with his mitt and finger.
% Pitcher get information of next pitch type/course and accept or reject it.
% Sometimes, their coach in their dugout sends his signal to catcher in advance.

% 配球の有効性
% Does the ability of devising a pitching strategy exist?
% This is quite interesting, but difficult question to answer in baseball analytics.
While a lot of stats are widely used to evaluate a player's performance, 
there are many unclear points around the pitching strategy.
The pitching strategy here means a strategy of battery (pitcher and catcher) to defeat the opponent batter.
Battery combines various pitch types, courses, and pitch combinations (e.g., inside fastball after outside breaking ball).
% During a game, suggesting the next pitch type/course is a catcher's task.
\textcolor{black}{During a game, a catcher often suggests where/which to pitch the next ball to a pitcher (especially in Japanese Professional Baseball (NPB)).}
The catcher then sends a signal to the pitcher with their mitt and finger.
The pitcher receives information about the next pitch type/course and accepts or rejects it.
Sometimes, their coach in their dugout sends the signal to the catcher in advance.

It is unknown whether the ability for selecting an appropriate pitching strategy exists.
\citet{woolner02} concluded the ability might exist, but cannot detect from batting results due to noise.
% , and many baseball fans and analysts do not seem to believe its existence.
\textcolor{black}{In other words, the ability does not include noises or lucks and an unmeasured variable.}
There are some quantitative studies on the effects of pitching sequences.
\citet{gray02} revealed that ``fastball after a series of late balls is effective'' by virtual experiments.
However, there are no statistics to evaluate the overall \textcolor{black}{effect} to select a pitching strategy.
We consider that two elements make it difficult to measure the effect of pitching strategy:
1) most public datasets do not include location data where the catcher demands a ball, 
2) there are a lot of confounders associated with pitching/batting results.

\begin{figure}[t]
  \centering
  \includegraphics[width=8.4cm]{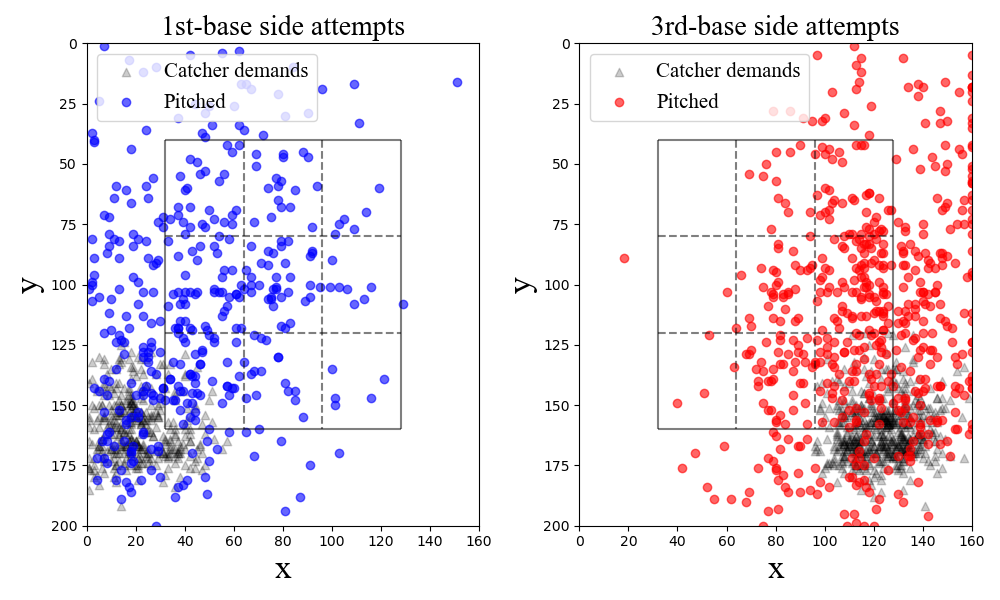}
  \caption{All-pitched location of four-seam where the catcher demands lower 1st-base/3rd-base side (regardless of batter's hand) from pitcher's perspective pitched by Taisuke Yamaoka, right-handed pitcher belongs to Orix Buffaloes in 2019 ($N = 935$ in total). Units are centimeters.}
  \label{fig: pitch_distribution}
\end{figure}

% 制球の困難さ
First, we explain why we need the location data where the catcher demands.
It is known that a pitcher cannot always pitch to the location of the catcher's demand perfectly.
For example, \citet{shinya17} revealed that pitched location follows a two-dimensional normal distribution.
Figure \ref{fig: pitch_distribution} shows the pitch distribution where the catcher demands a lower 1st-base/3rd-base side (unless otherwise noted, figures are from the pitcher's perspective).
There is a gap between demanded location and the actually pitched one.
Thus, a catcher needs to suggest the next pitch location considering the gap between the required course and the pitched one.
For the above reason, in the fielding team's position, we should measure the effects of pitching ``attempts'' to a specific course, not pitched.

In addition to the difficulty of observing the catcher's demand, 
many confounders associated with pitching/batting results make it difficult to evaluate pitching strategy.
For example, it is known that a good batter is more frequently pitched inside (especially in NPB).
The correlation between weighted On Base Average (wOBA; high wOBA means good batter), and the ratio of inside pitching attempts was $r = 0.31$ in our computation\footnote{We computed the correlation (Pearson's $r$-value) for the batters which have more than $300$ plate appearances in a season from 2014 to 2019 in NPB games. Please refer to Section \ref{ssec:dataset} for the definition of inside pitching attempts, and the detail.}.
If simply comparing the results where demanding inside with the outside,
the effect of outside pitching attempts can be overestimated 
since poor hitters are attempted outside more frequently than hard hitters.
Thus, we cannot compare them directly.

For the above reason, we propose a method of estimating the effect of demanding inside/outside with causal inference called stratified analysis using propensity score (e.g., \cite{rubin1997estimating,damluji2019percutaneous}), which can statistically \textcolor{black}{reduce} confounders and estimate the effects.
We define the effect of each demand as a variation of run expectancy, computed by pitching/batting results.
Thus, we develop the model to \textcolor{black}{compute} the propensity score (i.e., the probability of demanding inside) with confounding variables (e.g., the ratio of pitched inside for the batter, the pitcher in the season), and make up for unobserved data with it.
We then estimate the expected effect of demanding inside and outside.
% Furthermore, to reveal the effect by propensity score, 
% we group the data into 4 groups and estimated the effect for each layer. 
% 強打者には内角、内角がそもそも多い投手/打者への内角投球企画は有効であるはず
Furthermore, we investigate specific questions about unknown problems in baseball. 
% validate if the current operations are effective.
We group the data into some groups based on the \textcolor{black}{estimated} batter's batting ability, and the ratio of pitched inside for pitcher/batter, and estimated the effect for each group (i.e., stratified analysis).
For the \textcolor{black}{estimated} batter's batting ability, we examine the effectiveness of the inside pitch attempt against a good batter, because it is considered that a good batter is more frequently pitched inside (especially in NPB).
In the same way, we investigate the effect of the ratio of pitched inside for pitcher/batter because the pitcher would think their inside pitch attempts are effective and pitch more frequently.
% We reveal that outside pitching attempts were more effective than inside on average.
% Besides, the stratified analysis shows that inside pitching attempts were effective in the actual situation expected to demand an inside pitch.
% Besides, the stratified analysis shows that the outside pitching attempt is always effective despite of magnitude of the batter's batting ability, and the ratio of pitched inside for pitcher/batter.

In summary, the main contributions of the study are as follows: 
(1) We analyzed the effects of pitching attempts to a specific location \textcolor{black}{ (inside/outside)} for the first time;
(2) We used the stratified analysis using the propensity score to estimate the effect by removing confounders in complicated scenes of baseball;
(3) We revealed the effect of pitching attempts on a specific course exists.
The remainder of the paper is organized as follows. 
We review the related work in Section \ref{sec:related}. 
In Section \ref{sec:method}, we describe the proposed method.
Section \ref{sec:results} presents the experimental results. 
Finally, we discuss the results and present the conclusions in Sections \ref{sec:discussion} and \ref{sec:conclusion}, respectively. 

%%%%%%%%%%%%%%%%%%%%%%%%%%
\section{Related work}
\label{sec:related}
\vspace{-0pt}
A lot of formulas and stats were developed, and employed to evaluate \textcolor{black}{past} or predict \textcolor{black}{future} player's performance from a long time ago \citep{silver03, james10, lewis04, click06, tango07, beneventano12, costa12}.
These stats are computed by batting results.
One of the most essential perspectives when evaluating the player's performance is to consider uncertainty.
% In baseball, uncertainty always exists and that makes it difficult to estimate a player's true ability.
\textcolor{black}{Even if two players have the same ability, the two results can be different
due to uncertainty}.
To deal with the problem, some stats focus on the ``true ability'': strikeout, home run, walk \citep{davies10}.
These stats are considered to be stable for each player for evaluating the player's performance.

Recently, measurement technology and data platform have been developed, 
and all sensing data measured in MLB games are provided and accessible.
\textcolor{black}{It enables us to develop stats to evaluate player's performance using speed and angle of batted balls, such as expected weighted On Base Average (xwOBA) and expected Earned Run Average (xERA) \citep{xwoba, xera}.}
The speed and angle of batted balls are \textcolor{black}{relatively} stable stats for an individual player \textcolor{black}{(i.e., the value does not fluctuate significantly across seasons.).
These new stats can extract a player's potential  more validly.
Comparing with evaluating batting/pitching performance}, 
evaluating pitching strategy is a challenging field.
Some researches predicted the next pitch type using various machine learning methods \citep{hoang15, bock15}.
Besides, the methods for predicting the batting result are suggested \citep{martin19, harrison19, healey15}.
When \textcolor{black}{estimating} the ability to select a pitching strategy for the catcher, the simplest idea is computing ERA where the catcher is in the games, called catcher's ERA \citep{herrlin2015}.
\citet{woolner02} showed that the catcher's ERA computed using batting results cannot detect the difference in abilities.
It seems difficult to \textcolor{black}{estimate} the catcher's ability to select a pitching strategy with batting results.
Besides, there is no study with the data of catcher's demanding as far as we know.

\textcolor{black}{In team sports, propensity score matching was used to investigate the causal effect of some plays or timeouts in many sports such as going for the touchdown in American football \citep{yam2019lost}, clearing the puck in ice hockey \citep{toumi2019grapes}, the effectiveness of timeouts in basketball  \citep{gibbs2020causal}, and crossing the ball in soccer \citep{wu2021contextual}.
In the dynamic setting, \cite{vock2018estimating} applied a g-computation method to examine the effect of taking a pitch during a 3-0 count in MLB and \cite{fujii2022estimating} proposed a framework for estimating individual treatment effect (ITE) in basketball motions.
Among approaches without causal inference methods, in baseball, counterfactual predictions could be performed such as using simulation and supervised learning in team hitting strategies \citep{nakahara2022estimating} and using game theory and semi-supervised learning in a third base coach's decision making \citep{nakahara2022evaluating}.
Compared with these approaches, since the outcome prediction in our task was very difficult (in a more general pitching case than the study of \citep{vock2018estimating}), we used stratified analysis (not to predict outcomes) rather than regression analysis to predict outcomes.
}

\section{Methods}
\label{sec:method}

In this section, we describe the proposed method to clarify the effect of pitching attempts to specific locations. 
First, we describe the stratified analysis using the propensity score to estimate the effect of demanding inside/outside pitch.
Second, we introduce the new dataset, including the location where the catcher demands.
Finally, we describe the processing method to compute the propensity score.

\vspace{-0pt}
\subsection{Stratified analysis using propensity score}
% \label{ssec:stratified}
\vspace{-0pt}

Our motivation is to reveal the effect of demanding inside/outside, and we need to \textcolor{black}{reduce} confounders for \textcolor{black}{valid} estimations.
To deal with confounding, we employed methods introduced by \citet{rosenbaum85}, defined as follows.
% which is currently called potential-outcome model and has been widely regarded as a key to understanding causal effects.
$X$ is a confounding variable.
$Z \in (0, 1)$ is a treatment allocation.
$Z = 0$ corresponds to nontreatment, and $Z = 1$ corresponds to given treatment.
In this study, $Z = 0$ ($Z = 1$) means catcher demands outside (inside) pitch.
$Y$ is a potential outcome.
$Y^{(1)}$ ($Y^{(0)}$) denotes the outcome if (not) treatments are applied, in this case, the outcome if a catcher demands inside (outside).
\textcolor{black}{The outcome is defined as the variation of run expectancy before and after the event (i.e., pitch result or batting result) ($\Delta RE$).
For details, see Section \ref{ssec:preprocess}.}
For simplicity, denote $X$ in the $i$-th unit (sample or pitch) as $X_i$.
The same is true for the other variables.
$\tau$ denotes the average treatment effect (ATE), and is described as follows:
\begin{eqnarray}
\tau = E[Y^{(1)}] - {E[Y^{(0)}]}.
\end{eqnarray}

% As well as the total effect of demanding inside/outside, we are also interested in the effects in specific condition.
The main problem in causal inference is that we cannot observe $Y_i^{1}$ and $Y_i^{0}$ simultaneously, known as ``the fundamental problem of causal inference'' \citep{holland86}.
To overcome the problem, various methods are proposed, 
and one of the popular fundamental ideas is comparing groups (i.e., sets of units), which have similar confounding variables.
\textcolor{black}{It should be noted that we used the stratified analysis (not to predict outcomes) rather than regression analysis to predict outcomes.
This is because the outcome prediction in our task was very difficult in our preliminary experiment.}
The propensity score is considered as the criteria for measuring the similarity between each unit.
The propensity score (here denoted by $P(X)$) is the probability of being assigned treatment computed using confounding variables.
We can consider the distributions of confounding variables are the same between treatment and nontreatment groups when comparing data with a similar propensity score, according to \citet{rosenbaum85}.
That is, comparing data with similar propensity scores \textcolor{black}{reduces} confounding. 
% This is called balancing property.
Note that here we assume three popular assumptions in causal inference: consistency, exchangeability, and positivity \citep{cole2008constructing,austin2015moving}.
Exchangeability (or ignorable treatment assignment) is the assumption of no unmeasured confounders that affect treatment selection and outcomes.
\textcolor{black}{However in the actual baseball data, we cannot say that there is no possibility of that.
We also discuss the limitation of this study from this viewpoint in Section \ref{sec:discussion}.}
Consistency means that a pitch's potential outcome under the treatment actually received is equal to the pitch's observed outcome.
Positivity is the assumption that all pitches have a non-zero probability of receiving each treatment.

We use the Inverse Probability Weighting (IPW), one of the practical causal inference methods using propensity scores \citep{robins00}.
In IPW, the propensity score is used to make up for unobserved outcomes.
It is the same with comparing units which have similar propensity scores, mathematically.
IPW estimates the treatment effect ($\tau$) as follows.
\begin{align}
\hat{\tau} &= {E[\hat{Y}^{(1)}]} - {E[\hat{Y}^{(0)}]}, \label{eq:ATE}\\
{E[\hat{Y}^{(1)}]} &= \left. \sum_{i=1}^{N} \frac{Y_i^{(1)}Z_i}{P(X_i)} \right/ \sum_{i=1}^{N}\frac{Z_i}{P(X_i)}, \label{eq:ATE2}\\
{E[\hat{Y}^{(0)}]} &=  \left. \sum_{i=1}^{N} \frac{Y_i^{(0)}(1-Z_i)}{1-P(X_i)} \right/ \sum_{i=1}^{N}\frac{1-Z_i}{1-P(X_i)}, \label{eq:ATE3}
\end{align}
where $\hat{\tau}, \hat{Y}^{(0)}$, and $\hat{Y}^{(1)}$ are expected values of $\tau, Y^{(0)} $, and $Y^{(1)}$, respectively.
As an example, we consider the units that satisfy $X=x, Z=1, P(Z=1|X=x)=0.3$.
Then, $P(Z=0|X=x)=0.7$, and $70\%$ of units with $X=x$ do not be assigned treatment.
We cannot observe these outcomes with treatment.
To obtain the unobserved outcomes counterfactually, each observed outcome is multiplied by $1/P(Z=1|X=x)$ (i.e., $1/0.3$).
Since we cannot observe the true propensity score, we need to develop a prediction model for its estimation.
Using random forest, propensity scores were estimated by modeling the associations of covariates with treatment (i.e., the random forest classifies whether the treatment exists or not).
The number of trees was $130$, and the maximum depth of the tree was $9$.
% We used the random forest classification provided by scikit-learn, and the hyper-parameters were as follows:
% the number of decision trees were $130$, the maximum depth of the tree was $9$.

% 傾向スコアの分布は、PCIグループと非PCIグループの間で類似しており（図1）、傾向スコアの全範囲で交絡の制御が可能であることを示しています。
% As a verification of the causal inference, propensity score distributions were similar between treatment groups and nontreatment groups (Figure 1).
% Each covariates is ajusted adequately by IPW (Figure 2).
% These results show that control for confounding is possible across the entire range of the propensity score.

% 標準偏差
We compute the expected effect and confidence interval to examine the treatment effect to be statistically significant.
We compute this on average for ATE, and this on each interested group (e.g., the \textcolor{black}{estimated} batter's batting ability and the ratio of pitched inside for pitcher/batter) as a stratified analysis.
We employed the bootstrap method, which is one of the Monte Carlo methods, resampling data randomly to estimate the confidence interval (such as used in \cite{bock2012hitting}).
The bootstrap method repeats the following operation multiple times: 
extracting the same number of data as the original from the original dataset randomly \textcolor{black}{with replacement}, and estimating the effect using IPW.
Among various methods to estimate confidence interval of the treatment effect (e.g., \cite{haukoos2005advanced}), we used the normal approximation method in this paper.
It approximates the resampling distribution as the normal distribution.
Then, the confidence interval is calculated using the sample average and its standard deviation.
% The confidence interval is calculated using the standard normal distribution. 
% As the number of iterations increases, the distribution converges to a normal distribution according to the central limit theorem.
% Based on the obtained distribution, we can compute the confidence interval.

\vspace{-0pt}
\subsection{Dataset}
\label{ssec:dataset}
\vspace{-0pt}
% データの提供元とその理由
We used the dataset of NPB games held in 2014-2019 provided by Delta Inc. 
since it contains location data where the catcher demands.
Although most baseball studies used MLB data \citep{koseler17}, no MLB dataset includes the demanding location in public.
We used the pitch-by-pitch data (e.g., game situation, the location of a pitch, the location of catcher's demanding by catcher just before pitching, pitch type, pitch speed, and pitching/batting result) of  $1{,}591{,}586$ pitches.
We assume that demanding location indicates the catcher's intent since 
the pitch distribution of inside and outside are different and scattered around the edge of the inside and outside, respectively (Figure \ref{fig: pitch_distribution}).
Note that the location data is inputted manually (most teams collect sensing data themselves, but not in public in Japanese professional baseball).
We show the detail of features in Section \ref{ssec:preprocess}.

\begin{figure}[t]
  \centering
  \includegraphics[width=6cm]{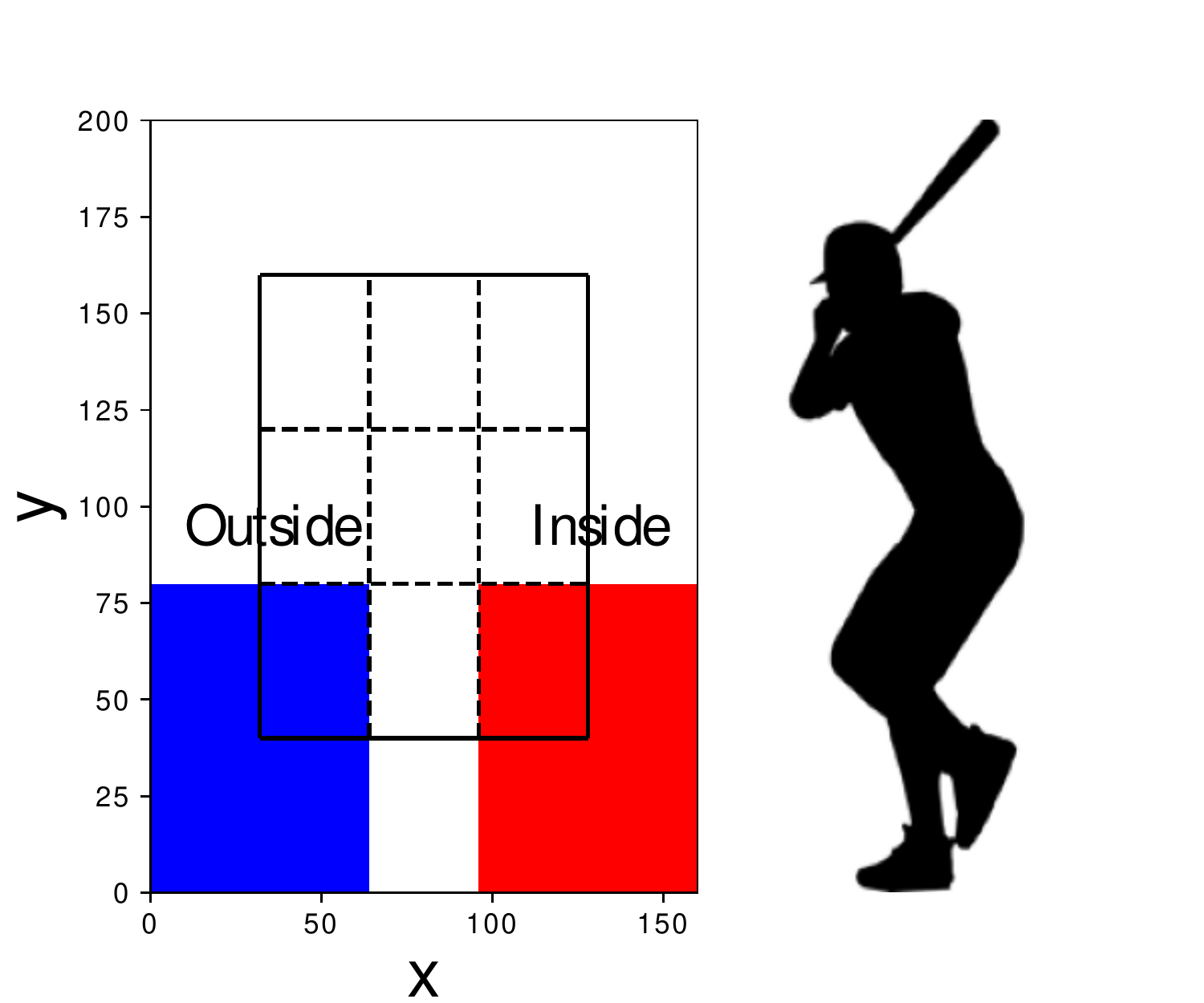}
  \caption{\textcolor{black}{The coordinate system of pitching location and the definition of inside/outside.} Units are centimeters.}
  \label{fig:in_out}
  
\end{figure}

% 構え座標の説明
% insideとoutsideの定義
\textcolor{black}{Figure \ref{fig:in_out} shows the coordinate system of pitching location and the definition of inside/outside pitch.}
Demanding and pitched locations are plotted on $160$ $\times$ $200$ px ($100 \times 125$ cm) campus.
The coordinate system has the $X$-axis horizontal and the $Y$-axis vertical with the origin $(0, 0)$ at the \textcolor{black}{lower} 1st-base side corner.
% We illustrate the definition of ``inside'' and ``outside'' in Figure \ref{fig:in_out}.
Note that we use a case when a pitcher faces a right-handed batter for a simple explanation (in the case of left-handed batter, reverse right-handed case).
We define ``the catcher demands inside (outside)'' where the plot is located in 
1) more right (left) than one-third of a home base's right side (left side),
and 2) lower than the top-third of the strike zone.
\textcolor{black}{The data with the high demanded ball was excluded due to the small sample size, and the data where the balls touched the ground was included.}
We group into ``the catcher demands inside'' where $D_x \geq 96$ and \textcolor{black}{$D_y \leq 80$}, 
``the catcher demands outside'' where $D_x \leq 64$ and \textcolor{black}{$D_y \leq 80$}.
Here, $D_x, D_y$ denotes the coordinates of the location of demand.

% 手で入力してます
% In our dataset, plot data (where catcher call for) is inputted manually with game videos.
% It is difficult to input plot data perfectly and it might contains error.

\vspace{-0pt}
\subsection{Data processing}
\label{ssec:preprocess}
\vspace{-0pt}

% 前処理
In this study, we used the data that games were held from March to June each season to compute feature values (e.g., batter's batting stats, the ratio of pitching inside for the pitcher, and so on).
Thus, we predicted them for the games held from July to September in each season.
% 4seamだけに限定,内角でも外角でもないものを除外
We used only data pitched four-seam fastball since the location of each breaking ball's demanding is fixed  
% almost all pitch types, except fastball, have only one demanding location
(e.g., most right-handed pitchers try pitching their slider into the left side of the home base from the pitcher's perspective).
The data on the batter's plate appearance or the pitcher's total faced batters less than $100$ through a season \textcolor{black}{were excluded due to the small samples.
In addition, the data grouped into neither inside nor outside were also excluded.}
We used $286{,}430$ pitches to execute causal inference.
The number of inside pitches was $91{,}186$, while that of outside pitches was $195{,}224$.

% アウトカウントなどをcontinuousへ
% とりあえず省略

% 投球結果および打席結果は得点スケールへ変換
\textcolor{black}{Here we describe the computation of the input feature for propensity score and the difference in run expectancy as the outcome.}
The original dataset in baseball includes pitch results (called strike, swinging strike, foul, ball) and batting results (e.g., hit, strikeout, and walk).
To evaluate pitch results and batting results with unified criteria, we converted pitching/batting results into \textcolor{black}{the run scale}.
\textcolor{black}{It is because there is a strong correlation between a win and earned runs, and 
the pitching strategy should be evaluated using runs.}
% the variation of run expectancy ($\Delta RE$).
The linear weights (LWTS) is one of the most popular ideas for converting each \textcolor{black}{event (pitching/batting results)} into run scale and ideas for evaluating each result in sabermetrics \citep{thorn15}.
In LWTS, the run expectancy is computed for each game situation, considering out count, and runners on base regularly (e.g., the run expectancy of $0$ out without runner is $0.44$).
With the run expectancy for each game situation, we can compute the variation of run expectancy as follows.
$$\Delta RE = RE_{after} - RE_{before},$$
where $RE_{after}$ and $RE_{before}$ are the run expectancy of after/before event.
\textcolor{black}{In other words, the value of each event is regarded as the variation of run expectancy.}
To compute the variation of run expectancy, we compute the expected  variation of run expectancy for each event as follows:
$$\Delta RE(event) = \frac{1}{N} \sum_{i \in event}^N{\Delta RE_i},$$
where $\Delta RE(event)$ is the expected variation of run expectancy for the specific event, observed $N$ times.
Table \ref{tab:re} shows the specific $\Delta RE$ of each event.

% 得点期待値の差分を効果として定義
% An effect of each demanding location is computed using $RE$.
% We define difference between after pitch $RE$ and before one as effect.
% We call it variation of run expectancy($\Delta RE$) .
%Mathematically, it is computed as follows.
% $$\Delta RE = RE_{after pitch} - RE_{before pitch}$$

\begin{table*}[t]
  \begin{center}
    \caption{Variation of run expectancy ($\Delta RE$) for each event.}
    \begin{tabular}{|c|r||c|r|} \hline
      Result & \multicolumn{1}{c||}{$\Delta RE$} & Result & \multicolumn{1}{c|}{$\Delta RE$} \\ \hline
      strike & -0.038 & hit by pitch & 0.311 \\
      ball & 0.032 & intended walk & 0.155 \\
      single & 0.437 & bunt & -0.133 \\
      double & 0.786 & bunt and error & 0.720 \\
      triple & 1.117 & bunt strike out & -0.249 \\
      home run & 1.408 & bunt and fielder's choice & 0.700 \\
      field out & -0.235 & sacrifice fly & 0.007 \\
      double play & -0.746 & sacrifice fly and error & 0.713 \\
      foul fly & -0.266 &  error & 0.664\\
      swinging strike out & -0.255 & fielding interference & -0.337 \\
      called strike out & -0.238 & batting interference & 0.516 \\
      uncaught third strike & 0.294 & foul liner & -0.426\\
      walk & 0.292 & & \\ \hline
    \end{tabular}
    \label{tab:re}
  \end{center}
\end{table*}

% 自信度関数の導入
For \textcolor{black}{valid} computation of the propensity score, we propose another feature value called pitch confidence $C(n)$.
It indicates how pitcher and catcher are confident about inside/four-seam pitch in $n$-th pitch.
We only illustrate the inside confidence model.
The four-seam confidence function is the same as the inside one.
The formula is computed as follows:
\begin{align} \label{eq:Cin}
  C_{in}(n) = \begin{cases}
    \alpha\Delta RE(n - 1) +(1 - \alpha)C_{in}(n - 2) \\  ~~~~~~~~~~~~~~~~~~~~~\text{($(n-1)$th pitch was inside)} \\
    C_{in}(n-1)  ~~~ \text{(otherwise)},
  \end{cases}
\end{align}
\begin{align} \label{eq:Cout}
  C_{out}(n) = \begin{cases}
    \alpha\Delta RE(n - 1) +(1 - \alpha)C_{out}(n - 2) \\  ~~~~~~~~~~~~~~~~~~~~~\text{($(n-1)$th pitch was outside)} \\
    C_{out}(n-1)  ~~~ \text{(otherwise)},
  \end{cases}
\end{align}
\begin{align} \label{eq:Cn}
  C(n) = C_{in}(n) - C_{out}(n),
\end{align}
\if0
\[
  C_{in}(n) = \begin{cases}
    \alpha\Delta RE(n - 1) + (1 - \alpha)C_{in}(n - 2) & \text{($(n-1)$th pitch was in side)} \\
    C_{in}(n-1) & \text{(otherwise)}
  \end{cases}
\]
\[
  C_{out}(n) = \begin{cases}
    \alpha\Delta RE(n - 1) + (1 - \alpha)C_{out}(n - 2) & \text{($(n-1)$th pitch was out side)} \\
    C_{out}(n-1) & \text{(otherwise)}
  \end{cases}
\]
$$C(n) = C_{in}(n) - C_{out}(n)$$
\fi
%$s\Delta RE$ denotes amounts of variation of run expectancy.
where $\alpha$ denotes a constant value ($0 < \alpha < 1$).
It expresses how much ``confidence'' remembers recent results.
The larger, the more ``confidence'' forgets old pitch results and weights recent pitch results.
The smaller, the more ``confidence'' remembers old results well and does not weight recent results relatively.
In this study, we added two confidence models. One is $\alpha$ = 0.6 and the other is $\alpha = 0.001$, determined empirically.

Finally, we used $18$ feature values for prediction, as follows:
game condition (ball count, out count, runner on base, run difference, whether the batter's hand is the same as the pitcher's hand or not), 
pitcher (total pitch in the game, pitch result one pitch ago, pitch result two-pitch ago, pitch speed one pitch ago, pitch speed two-pitch ago, the confidence of pitching into the inside ($\alpha=0.6, 0.01$), the confidence of pitching four-seam ($\alpha=0.6, 0.01$), previous batting result, the ratio of pitched into inside in the season), 
batter (the ratio of pitched into inside in the season, wOBA).

\vspace{-0pt}
\section{Results}
\label{sec:results}
\vspace{-0pt}
In this section, we first verified the propensity scores, which is the basis of the causal inference method.
Second, we show the contribution of each variable to propensity score computation.
Third, we show the estimated total treatment effect (i.e., the effect of demanding inside/outside).
Finally, we show the treatment effect within the groups based on a) the magnitude of the \textcolor{black}{estimated} batter's batting ability (wOBA \textcolor{black}{in this case}), 
b) that of the ratio of the inside pitch for the batter, c) that of the ratio of the inside pitch for the pitcher (i.e., stratified analysis).
% It consists of four parts: 1) estimating the total average effect and estimating the average effects within the groups based on the batter's batting ability (wOBA), the ratio of inside pitch for batter/pitcher, respectively (i.e., stratified analysis using propensity scores).

\vspace{-0pt}
\subsection{Verification of propensity scores}
\label{ssec:verifyPS}
\vspace{-0pt}

\begin{figure*}[t]
  \centering
  \includegraphics[width=11.6cm]{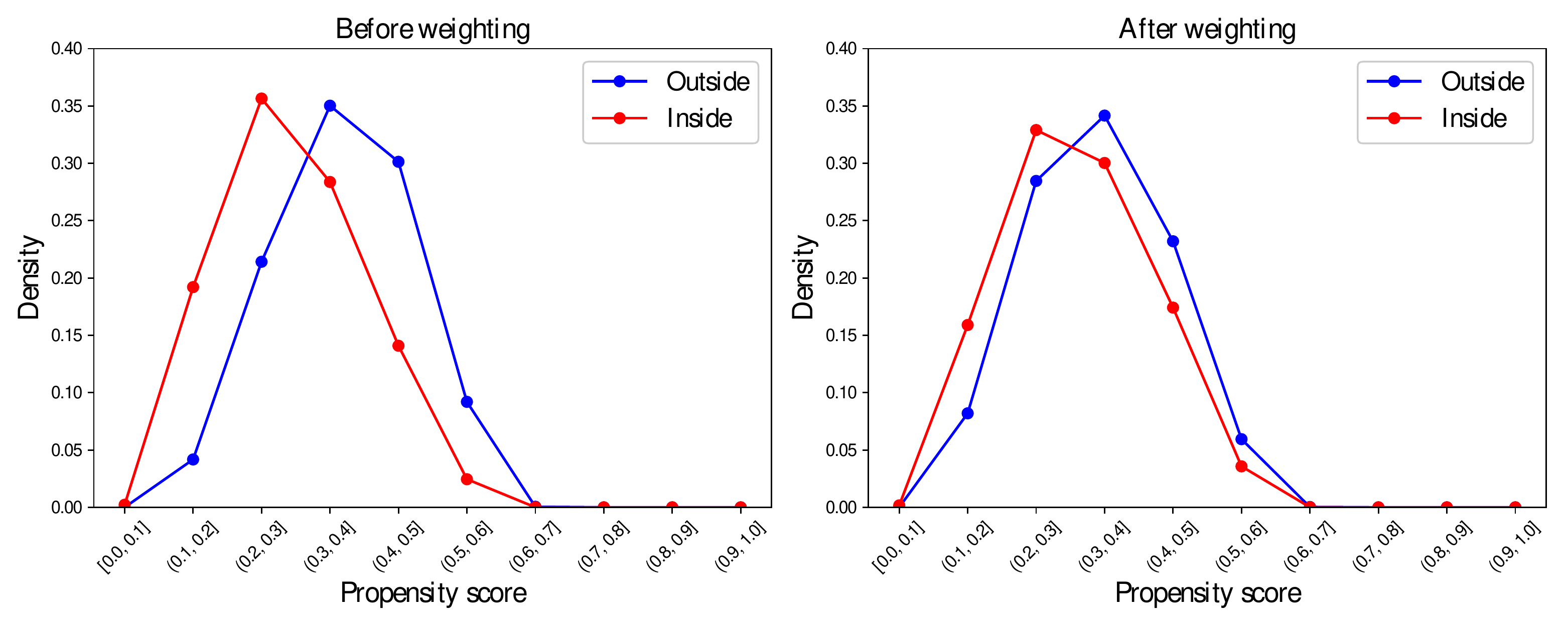}
  \caption{Overlap plot of the density of the propensity scores in (left) raw data and (right) weighted data via Inverse Probability Weighting (IPW). Although there was a large difference in the distribution of demanding outside and inside (left), the weighted results in outside and inside were similar (right).}
  \label{fig:psdis}
  
\end{figure*}

\begin{figure}[t]
  \centering
  \includegraphics[width=11cm]{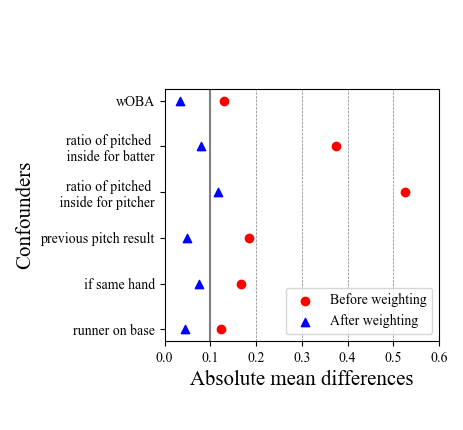}
  \caption{Absolute Standardized Absolute Mean distance (ASAM). Red and blue markers indicate before/after adjusted ASAM, respectively. Each ASAM should be lower than $0.1$.}
  \label{fig:asam}
\end{figure}

% 傾向スコアの分布は、PCIグループと非PCIグループの間で類似しており（図1）、傾向スコアの全範囲で交絡の制御が可能であることを示しています。
As a verification of the causal inference, propensity score distributions must be similar between treatment and nontreatment groups.
\textcolor{black}{Figure \ref{fig:psdis} shows the plots of the density of the propensity scores in (left) raw data and (right) weighted data via IPW.}
We confirmed that the distribution was adjusted by the IPW method using the propensity score adequately (for details, see Figure \ref{fig:psdis} and the caption).

Next, each covariate must be balanced between the treatment and nontreatment groups.
We employed Average Standardized Absolute Mean distance (ASAM) to balance them, and it is computed as follows.
\begin{eqnarray}
ASAM &=& \frac{|\overline{x_{in}} - \overline{x_{out}}|}{s}, \\
s &=& \sqrt{\frac{n_{in} s_{in}^2 + n_{out} s_{out}^2}{n_{in} + n_{out}}},
\end{eqnarray}
where $x_{in}$ and $x_{out}$ are the confounders with pitched inside and outside, respectively.
$\overline{x_{in}}$ and $\overline{x_{out}}$ are the arithmetical means of $x_{in}$ and $x_{out}$.
$n_{in}$ and $n_{out}$ are the number of data with pitched inside and outside, respectively.
$s_{in}$ and $s_{out}$ are the standard deviation of $x_{in}$ and $x_{out}$, respectively.
According to the previous work \citep{cannas2019comparison}, researchers should aim at obtaining an ASAM lower than 0.1 for as many variables as possible.
% We set criteria as $ASAM \leq 0.1$ to consider that they are sufficiently balanced to execute causal inference \citep{mccaffrey04}.
Figure \ref{fig:asam} shows the adjusted and nonadjusted ASAM.
% , where nonadjusted ASAM is greater than $0.1$.
Although there was only one covariate (i.e., the ratio of pitched inside for pitcher) over $0.1$, the value $0.116$ was close to $0.1$.
We consider each covariate was adjusted adequately using a propensity score since almost all of ASAM of confounding variables were lower than $0.1$.
% We confirmed that they are well-balanced.
These results show that control for confounding would be possible across the entire range of the propensity score.

\subsection{Contribution of each variable}
\label{ssec:shap}

\begin{figure}[t]
  \centering
  \includegraphics[width=11cm]{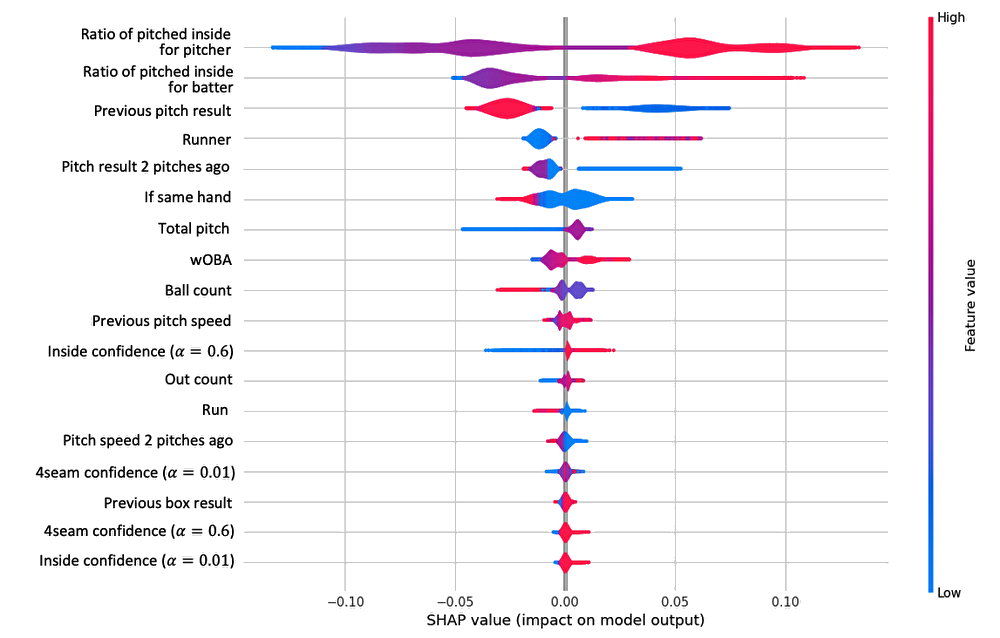}
  \caption{Contribution of the input variables to the propensity score \textcolor{black}{model} prediction. 
    The variables related to the prediction are shown in the order of their contribution.
    Each dot (but merged) represents each event. The color represents the value of the feature (blue and red indicate low and high, respectively). The horizontal axis shows the impact on the prediction (strongly positive and negative impacts are plotted to the right and left, respectively). }
  \label{fig:shap}
\end{figure}

The contribution of the input variables to the propensity score \textcolor{black}{model} prediction was computed by SHAP (SHapley Additive exPlanations) \citep{Lundberg17}, which utilizes an interpretable approximate model of the original nonlinear prediction model and also used for sports analysis \citep{toda2021evaluation}.
Figure \ref{fig:shap} shows the contribution of the input variables (SHAP).
Note that variables including ``result'' are converted into run scale, and a negative value means a pitcher is in advantage.
The ratios of pitched inside in the season for pitcher/batter were important for prediction.
Following them, previous pitch result, runner, whether pitcher's hand is the same as the batter's one were also important.

% permutation importanceの話
% Next, we computed the permutation importance to clarify where a inside pitching was attempted \citep{altmann10permutation}.
% Permutation importance is one of useful measurements to compute the importance of each feature value.
% It computes the importance of a feature value by comparing the ordinal prediction accuracy with the one where the target feature value is reordered randomly.
% Table \ref{tab:importance} shows the result.
% The ratio of pitched inside in the season for pitcher/batter are the most important for prediction.
% Following them, previous pitch result, pitcher/batter's hand are also important.

\subsection{Total ATE}
\label{ssec:total}

We then analyzed the total ATE computed in Equation (\ref{eq:ATE}).
The positive effect means the outside pitching attempt is effective, in contrast,
the negative effect means the inside pitching attempt is effective.
%\textcolor{black}{The ATE directly comparing the average outcomes between outside pitches and inside ones (not causal inference ) was $7.71 \times 10^{-3}$.}
\textcolor{black}{The ATE was $6.29 \times 10 ^ {-3}$, and the confidence interval ($99\%$) was $6.21 \times 10 ^ {-3} \leq ATE \leq 6.36 \times 10 ^ {-3}$.}
It means that demanding an outside pitch is more effective than demanding an inside pitch.
Since the effect of getting a strike is about $-3.8 \times 10^{-2}$,
it did not seem a large effect.

\textcolor{black}{Next, for comparison, we just compared the variation of run expectancy between inside attempts and outside ones (without IPW).
Results show that that of an outside pitching attempt was $7.71 \times 10^{-3}$ runs/pitch smaller than that of an inside one.
The results suggest that the outside pitching attempt was more effective than the inside pitching attempt compared with the analysis with IPW, which overestimated the value of outside pitching attempts.}

\vspace{-0pt}
\subsection{Stratified analysis}
\label{ssec:stratified}
\begin{figure}[t]
  \centering
  \includegraphics[width=11.6cm]{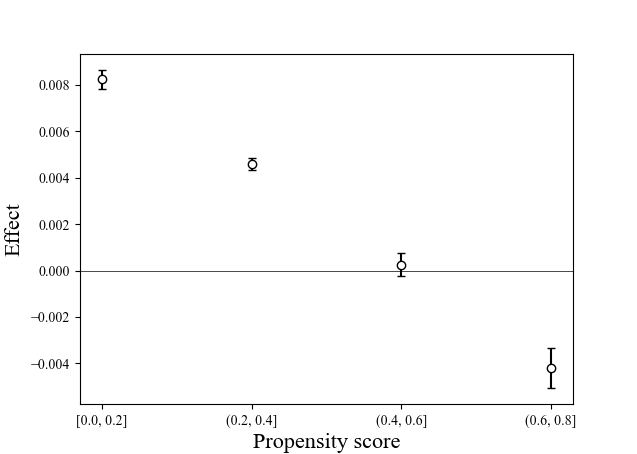}
  \caption{The estimated treatment effects for each group based on (a) \textcolor{black}{estimated batting ability (wOBA)}, (b) ratio of inside pitch for batter, (c) ratio of inside pitch for pitcher. 
  The vertical axis is the estimated treatment effects via IPW, in which the scale is run/pitch. We eliminated the data where the data with the confounding variable was greater than the criterion, due to the small sample size.}
  \label{fig:effect}
  \vspace{-0pt}
\end{figure}

\vspace{-0pt}
Next, we show the treatment effect within the groups based on a) the magnitude of the \textcolor{black}{estimated} batter's batting ability (wOBA \textcolor{black}{in this case}), 
b) that of the ratio of the inside pitch for the batter, c) that of the ratio of the inside pitch for the pitcher (Figure \ref{fig:effect}).
Each data was divided into \textcolor{black}{some} groups based on the magnitude of each confounding variable.
We eliminated the data with the confounding variable greater than the \textcolor{black}{$10,000$ samples}.
% Next, we show the treatment effect for each group split by propensity score (Figure \ref{fig:effect}).
% Each data was divided into four groups based on the magnitude of the propensity score.
% We eliminated the data with a its propensity score greater than $0.8$, due to the small sample size.
Each point represents a mean value, and the extended line represents a ($99\%$) confidence interval computed using the bootstrap method.
The result indicates that the pitching attempt to the outside is effective, regardless of the magnitude of each confounding variable.

\textcolor{black}{For the wOBA levels in Fig. 6(a), comparing groups with wOBA less than 0.3 and 0.3 or more, the effectiveness of pitching to the outside decreased when wOBA was large.
For the ratio of the inside pitch for the batter in Fig. 6(b), an outside pitch attempt was effective even against batters frequently pitched inside ($0.4$ or more), but these effects were smaller than the rest of the groups.
For the ratio of the inside pitch for the pitcher in Fig. 6(c), an outside pitch attempt was also effective even for pitchers frequently pitched inside ($0.6$ or more), but these effects were smaller than the rest of the groups.}

% Table \ref{tab:con_pro} shows the expected value of counfounding variables split by propensity scores.
% "is same hand" is boolean variable, $1$ represents the pitcher's hand and the batter's hand are same, and $0$ represents that are different.

% \begin{table*}[ht]
%   \begin{center}
%     \caption{the expected value of confounding variables split by propensity scores}
%     \begin{tabular}{|c|r|r|r|r|r|r|r|} \hline
%       propensity score & \multicolumn{1}{c|}{wOBA} & \multicolumn{1}{c|}{out count} & \multicolumn{1}{c|}{ball count} & \multicolumn{1}{c|}{strike count} &  \multicolumn{1}{c|}{is same hand} \\ \hline
%       $[0.0, 0.2]$ & .314 & 0.87 & 0.96 & 0.52 & 0.66 \\ 
%       $(0.2, 0.4]$ & .324 & 1.01 & 0.92 & 0.83 & 0.47 \\
%       $(0.4, 0.6]$ & .331 & 1.04 & 1.04 & 1.14 & 0.33 \\
%       $(0.6, 0.8]$ & .331 & 0.96 & 1.14 & 1.33 & 0.22 \\ \hline
%     \end{tabular}
%     \label{tab:con_pro}
%   \end{center}
% \end{table*}

%%%%%%%%%%%%%%%%%%%%%%%%%%%%%%%
\vspace{-0pt}
\section{Discussion}
\label{sec:discussion} 
\vspace*{-0pt}
In this study, we clarified the effect of pitching attempts to a specific location (\textcolor{black}{e.g}., inside or outside) in baseball.
In this section, we discuss the results of ATE, stratified analysis, propensity scores, and limitations and future perspectives of this study. 
% We employed stratified analysis using confounding variables to evaluate the effects using the dataset including location data where the catcher demands a ball.
% 外角への投球企画の方が有効
As shown in Section \ref{ssec:total}, the pitching attempt to the outside was found to be more effective than to the inside when pitching the four-seam fastball.
In NPB, the pitching strategy is based on pitching to the outside with an occasional mix of pitching to the inside, generally.
An inside pitching attempt is considered risky and requires more courage to pitch compared to an outside one.
This common belief is consistent with the results.

% 捕手評価につながる話
The results revealed that the total ATE of \textcolor{black}{selecting outside pitch} was about \textcolor{black}{$6.29 \times 10^{-3}$} runs.
\textcolor{black}{We assume that every straight pitch has a right pitching strategy and the effect is about $6.29 \times 10^{-3}$ (i.e., a right pitching strategy reduces  $6.29 \times 10^{-3}$ runs/pitch than the wrong one).
Then, we also assume that $150$ pitches are thrown in a game and $50\%$ of that are four-seam fastball ($75$ four-seam fastballs are thrown in a game).
Under these assumptions, the maximum contribution of saving runs by selecting an appropriate pitching course was estimated at $0.47$ runs/game.}
Although the difference of pitching strategy among catchers needs to be considered,
this implies that \textcolor{black}{we can measure the effect of }selecting pitch course \textcolor{black}{to save runs} (and type) may exist.

% 層別分析の解釈
% それぞれの共変量によらず外角が有効でした
% （被）内角企画が多い→内角が有効だと思って投げている→それでも外角が有効
% 今回のモデルでは表現しきれていないかもだけど、内角が有効な場面は想像以上に限られるのかもしれない
As shown in Section \ref{ssec:stratified}, the outside pitching attempt was found to be always more effective in each situation.
For \textcolor{black}{the estmiated} batting ability, it is known that a hard hitter is more frequently pitched inside (especially in NPB), 
which means they think an inside pitch attempt is effective against a good batter.
\textcolor{black}{
For the wOBA levels in Fig. \ref{fig:effect}(a), the effectiveness of pitching to the outside decreased when wOBA was large.}
% (Figure \ref{fig:effect} (a)) shows that an outside pitch attempt was also effective against a good batter ($0.4 \leq wOBA < 0.45$).
% However, the estimated effect of inside pitching attempts against a good batter seems almost the same with not good batter, and less effective than outside one \textcolor{black}{(Figure \ref{fig:effect} (b))}.
% \textcolor{black}{
% Figure \ref{fig:effect} (b) shows the similar result.
% An outside pitch attempt was also effective even against batters frequently pitched inside ($0.4 \leq $ inside ratio $< 0.8$), but these effects were smaller than the rest of groups.}
\textcolor{black}{For the ratio of the inside pitch for the batter/pitcher in Figs. \ref{fig:effect}(b) and (c), an outside pitch attempt was effective even against batters and for pitchers frequently pitched inside, but these effects were smaller than the rest of the groups.
Although the pitch attempt to the inside was less effective than the pitch attempt to the outside from the perspective of the outside pitch attempt, overall, the pitch to the outside was more effective than those to the inside.
}
Considering the difficulty of controlling the location of the pitch, an inside pitching attempt might be overrated.
% The same is true for the ratio of inside pitching for batter/pitcher.
The scene where an inside pitching attempt \textcolor{black}{was effective} might be limited than expected.

% SHAPの話
We also computed the contribution of each variable to the propensity score \textcolor{black}{model} prediction (whether the catcher demands an inside pitch or not).
We clarified that the ratios of pitched inside for pitcher/batter were important, which seems to be obvious.
On the other hand, we also clarified that some variables contributed to predicting the propensity score \textcolor{black}{model} unexpectedly.
Previous pitch results one/two pitch ago were the third and fifth most important variables, respectively.
For both of them, the high SHAP value was associated with a low value, which means they tend to try to pitch inside more frequently where they are in advantage.
It might be also an important insight since there is no study about predicting the course of catcher's demand ever.

% 傾向スコアと内角への投球企画有効性
% The higher the propensity score, the more effective pitching attempt to the inside tends to be (Figure \ref{fig:effect}).
% Since the propensity score is an estimated probability of attempts to inside computed using covariates, 
% the pitching attempt to inside is effective in situations where the probability of attempts to the inside is high.
% This result implies that the catcher seems to adequately select the situation where the inside pitching attempt is effective  in practice. 
% wOBA, "is same hand" (which indicates whether the batter's hand is same with the pitcher's one) were found to have a strong correlation with the propensity score.
% Although we have to be careful that the associations is just correlation, not causal, that implies the catcher might consider "is same hand" and wOBA (batting ability) when choosing pitching course.

% 手法の問題点
If you are a baseball fan, you can easily understand that it is inappropriate to conclude that 
the pitcher should try to always pitch to the outside from these results.
It is because the ratio of pitching inside and outside needs to be balanced to keep their effects.
% This is because one factor that constitutes the effectiveness of pitching to the outside is 
% pitched to the inside, and pitching to inside is considered necessary to keep the effect of pitching to outside, more or less.
If all pitches were thrown to the outside, the batter could prepare for hitting the outside pitch
(e.g., standing closer to home base and stepping outside).
The batter only needs to focus on hitting the same course, improving the reproducibility of the \textcolor{black}{accurate} swing to hit a ball.
Thus, the effect of the scattering pitching course should be considered to obtain a more practical result.

From the perspective of causal inference, this study has mainly two limitations. 
First, there might be unmeasured confounders because we can measure and use a limited number of variables in baseball games.
\textcolor{black}{For example, batter's intentions and pitcher/batter's condition are possible unobserved confounders.}
Second, datasets of baseball games are longitudinal in each game and in all games for each team.
Although we added the history of performances for batters and pitchers to the input features, we need to consider the models including time-varying confounders and treatment for future work.

In this study, we employed the IPW method that estimates ATE, adjusting the distribution, and comparing by groups.
Estimating the effect by comparing groups, such as IPW, is a common and traditional idea in causal inference.
On the other hand, various causal inference methods have been proposed, which can estimate Individual Treatment Effect (ITE) (e.g., \cite{kunzel2019}).
% It means that estimating specific effect values for individual data.
These methods estimate the potential outcome, which is not observed in real, counterfactually.
While the IPW method reduces a causal problem to predict the probability of allocated treatment for each data, 
the method estimating ITE reduces it to predict counterfactual pitching/batting results.
It is equivalent to predicting results with game information.
Predicting pitching/batting result is more difficult compared to predicting treatment probability.
Although we did not employ a method for estimating ITE due to the difficulty in the prediction, we are sure that estimating ITE will extend or improve this study.

%%%%%%%%%%%%%%%%%%%%%%%%%%%%%%%
\vspace{-0pt}
\section{Conclusions}
\label{sec:conclusion} 
\vspace*{-0pt}

In this study, a causal inference method using the location of the catcher mitt right before pitching revealed that an outside pitching attempt is more effective than an inside pitching attempt.
We also found that the outside pitching attempt was always effective despite the magnitude of the ratio of pitched inside for pitcher/batter.
Besides, we revealed that the previous pitch result is also important to predict if the catcher demands an inside pitch.
% We also found that the inside pitching attempt is effective in the situation where the catcher often demands an inside pitch, and indicating that the catcher's pitching strategy of selecting a course makes sense.
% The probability of demanding a inside pitch (propensity score) are strongly associated with wOBA, "is same hand" (which indicates whether the batter's hand is same with the pitcher's one).
% These results indicates that the effectiveness of choosing pitching course considering the gap between demanded pitching course and actual pitched course.
% It is quite useful knowledge when selecting a pitching strategy to minimize allowed runs in practice.

For future perspectives, the application of this study is 1) evaluating the catcher's selection of pitching strategy and 2) seeking a better pitching strategy.
% This research indicated that choosing a pitching course has run values.
% It implies the ability to save allowed runs by choosing pitch course (type) may exist.
First, our approach can reveal where courses and which types to pitch is effective in each game situation.
This enables us to evaluate each pitching strategy.
If we can compute ITE in our problem, we can evaluate each catcher's selections of pitching strategy.

Second, estimating the effect of each pitch makes it possible to seek a better pitching strategy, if we can develop a pitching strategy model.
Some Japanese companies are developing an automatic selecting pitching strategy system called ``AI catcher''.
Systems for predicting better strategy can entertain enthusiastic baseball fans as well as support decision-making.

\section*{Acknowledgments}
\textcolor{black}{The data was provided by Delta Inc. }
This work was supported by JSPS KAKENHI (Grant Numbers 19H04941 and 20H04075) and JST PRESTO (JPMJPR20CA).

\ifarxiv

\else
\bibliographystyle{DeGruyter}
\bibliography{main}
\fi
\end{document}